\documentclass[twoside,draft]{article}
\usepackage{latexsym,e-journal}

\usepackage[letterpaper]{geometry}
\usepackage[utf8]{inputenc}
\usepackage[T1]{fontenc}
\usepackage[english]{babel}

\usepackage{amssymb,amsfonts,amsmath}

\usepackage[perpage,symbol*]{footmisc}
\usepackage[final]{graphicx}
\usepackage{pstricks}
\usepackage{cite}

\usepackage[varg]{txfonts}

\begin{document}

\renewcommand{\refname}{References}
\renewcommand{\tablename}{\small Table}
\renewcommand{\figurename}{\small Fig.}
\renewcommand{\contentsname}{Contents}

\def \pteptitle {The Unpublished Feynman Diagram IIc}
\def \ptepauthor {Oliver Consa}

\twocolumn[%

\begin{center}
\renewcommand{\baselinestretch}{0.93}
{\Large\bfseries \pteptitle
}\par
\renewcommand{\baselinestretch}{1.0}
\bigskip
Oliver Consa\\ 
{\footnotesize  Department of Physics and Nuclear Engineering, Universitat Politècnica de Catalunya \\ 
Campus Nord, C. Jordi Girona, 1-3, 08034 Barcelona, Spain\rule{0pt}{8pt}\\
E-mail: oliver.consa@gmail.com
}\par
\medskip
{\small\parbox{11cm}{%
Quantum Electrodynamics (QED) is considered the most accurate theory in the history of science. However, this precision is limited to a single experimental value: the anomalous magnetic moment of the electron (g-factor). The calculation of the electron g-factor was carried out in 1950 by Karplus and Kroll. Seven years later, Petermann detected and corrected a serious error in the calculation of a Feynman diagram; however, neither the original calculation nor the subsequent correction was ever published. Therefore, the entire prestige of the QED depends on the calculation of a single Feynman diagram (IIc) that has never been published and cannot be independently verified.
}}
\smallskip
\end{center}] {%

\setcounter{section}{0}
\setcounter{equation}{0}
\setcounter{figure}{0}
\setcounter{table}{0}
\setcounter{page}{1}

\markboth{\ptepauthor. \pteptitle}{\ptepauthor. \pteptitle}

\markright{\ptepauthor. \pteptitle}
\section{Introduction}
\markright{\ptepauthor. \pteptitle}

According to the Dirac equation, the value of the magnetic moment of the electron should be exactly one Bohr magneton. In 1947 it was discovered that the experimental value of the magnetic moment of the electron presented an anomaly of 0.1\% with respect to the theoretical value \cite{Rabi}\cite{Breit}. This anomaly was called the electron g-factor.

\begin{equation} 
\mu_e =g \mu_B = g \frac{e\hbar}{2m_e}
\end{equation}

Schwinger carried out the first theoretical calculation of the electron g-factor obtaining a value very similar to the experimental value. This value is known as the Schwinger factor \cite{Schwinger}.

\begin{equation} 
g = 1 + \frac{\alpha}{2 \pi} = 1.001162 
\end{equation}

According to Quantum Electrodynamics (QED), the theoretical value of the electron g-factor is obtained by calculating the coefficients of a number series called the Dyson series \cite{Dyson}. When Feynman, Schwinger, and Tomonaga received the 1965 Nobel Prize for the development of QED, only the first two coefficients in the series had been calculated. The rest of the coefficients in the Dyson series were calculated many years later with the help of supercomputers.

\begin{equation}
g = C_1 \left(\frac{\alpha}{\pi}\right) + 
C_2 \left(\frac{\alpha}{\pi}\right)^2 + 
C_3 \left(\frac{\alpha}{\pi}\right)^3 + 
C_4 \left(\frac{\alpha}{\pi}\right)^4 +
C_5 \left(\frac{\alpha}{\pi}\right)^5 ...
\end{equation}

Each coefficient in the series requires the calculation of an increasing number of Feynman diagrams. The first coefficient in the Dyson series is the Schwinger factor and has an exact value of 0.5. The second coefficient was calculated in 1950 by Karplus and Kroll \cite{KK}, who obtained a result of -2.973. This result was corrected seven years later by Petermann \cite{Peter}, who obtained a result of -0.328, almost 10 times lower than the previous calculation. 

\begin{equation} 
g =  1 + \frac {1}{2} \left(\frac {\alpha}{ \pi}\right) - 0,328 \left(\frac {\alpha}{ \pi}\right)^2 = 1,0011596 
\end{equation}

The error was found in the calculation of the Feynman diagram  IIc. According to the Karplus and Kroll's, original calculation, the value of diagram IIc was -3.178 while in the Petermann correction the value of diagram IIc was -0.564.

\begin{figure}[ht]
	\centering
	\includegraphics[scale=0.6]{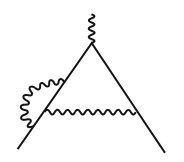}
	\vspace*{-3mm}
\caption{Feynman diagram IIc.}
\end{figure}

The entire prestige of the QED is based on its impressive level of precision of the electron g-factor. Currently the QED allows the achievement of the electron g-factor with a precision of 12 decimal places of the theoretical value with respect to the experimental value.
\begin{itemize}
\item 2008 Gabrielse's experimental value \cite{Gabrielse}:
\begin{itemize}
\item[$\ast$]  $1.001,159,652,180,73(28)$
\end{itemize}
\item 2018 Kinoshita's theoretical value \cite{Kinoshita}:
\begin{itemize}
\item[$\ast$] $1.001,159,652,182,032(720)$ 
\end{itemize}
\end{itemize}

The calculation of the electron g-factor is based on the calculation of the second coefficient of the Dyson Series. The second coefficient of the Dyson series is based on the calculation of the Feynman diagram IIc. Therefore, the calculation of the Feynman diagram IIc performed by Karplus and Kroll in 1950 \cite{KK} can be considered the most important calculation in the history of modern physics. 

Surprisingly, the original calculation of this diagram  IIc turned out to be completely wrong and was corrected seven years after its publication.  Inexplicably, both the original Feynman diagram IIc calculation and the subsequent correction have never been published, so the most important calculation in the history of modern physics cannot be independently verified.

\markright{\ptepauthor. \pteptitle}
\section{Original calculation}
\markright{\ptepauthor. \pteptitle}

\subsection{Karplus and Kroll's paper}

In 1949, Gardner and Purcell \cite{Gardner} published an new experimental result for the electron g-factor of 1.001146. In response, Karplus and Kroll performed the necessary calculations to obtain the second coefficient in the Dyson series.

In 1950, Karplus and Kroll \cite{KK} published a value of -2.973 for the second Dyson series coefficient and a new theoretical value of 1.001147 for the electron g-factor. In good agreement with the experimental data.

\begin{equation} 
g =  1 + \frac {\alpha}{2 \pi} - 2.973 \left(\frac {\alpha}{\pi}\right)^2 = 1.001147  
\end{equation}

The paper, published the February 14 in the Physical Review Journal 77, consists of 14 pages full of complex mathematical calculations.

On the second page of the document, the authors indicate that to obtain the coefficient it is necessary to calculate 18 Feynman diagrams grouped in five groups (I, II, III, IV and V). However, on pages 3 and 4, they argue that groups III, IV and V are not necessary. Therefore, it is only necessary to calculate seven Feynman diagrams, identified as I, IIa, IIb, IIc, IId, IIe, IIf. A lot of calculations are done between pages 4 and 11 that only serve to show that diagrams IIb and IIf are not necessary either. Therefore, it is only necessary to calculate five Feynman diagrams (I, IIa, IIc, IId, IIe).

\begin{figure}[ht]
\centering
\includegraphics[scale=0.45]{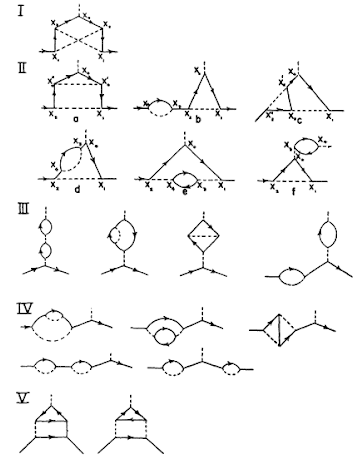}
\vspace*{-3mm}
\caption{Feynman Diagrams.}
\end{figure}

The calculation of diagrams IIe (0.016) and IId (-0.090) are performed on pages 11 and 12 respectively. It follows that \cite{KK}  \textbf{\textit{“The expressions for I, IIa and IIc become successively more complicated and very much more tedious to evaluate and cannot be given in detail here.”}}. In other words, the complete calculation of three of the five diagrams was never published. On page 13, the results of the three remaining diagrams are shown (I = -0.499, IIa = 0.778 and IIc = -3.178). Finally, page 14 of the paper presents the "Summary of Results" with the results of each of the five diagrams.

\begin{equation}
C_2 = I + IIa + IIc + IId + IIe = -2,973
\end{equation}

\renewcommand{\arraystretch}{1.3}
\begin{table}[ht]
\centering
\begin{tabular}{ |ccccc|c|} 
\hline
I & IIa & IIc & IId & IIe & Total \\
\hline
-0.499 & 0.778 & -3.178 & -0.090 & 0.016 & -2.973 \\
\hline
\end{tabular}
\caption{\label{tab:KK} Values of the five Feynman diagrams.}
\end{table}

From the analysis of the results it is evident that diagram IIc is the dominant diagram. Diagrams I and IIa are less relevant and practically cancel each other out. Diagrams IId and IIe are the only two diagrams whose calculations are included in the paper; however, their values are completely irrelevant.

The calculation of Feynman diagram IIc is made up of four components:
\begin{equation}
II_c = -\frac{323}{24} + \frac{31}{9} \pi ^2 -  \frac{49}{6} \pi ^2 ln (2) +  \frac{107}{4} \zeta(3) 
\end{equation}

\renewcommand{\arraystretch}{1.3}
\begin{table}[ht]
\centering
\begin{tabular}{ |cccc|c|} 
\hline
Constant & $\pi^2$ & $\pi^2ln2$ & $\zeta(3)$ & Total  \\
\hline
-13,458 & 33,995 & -55,868 & 32,153 & -3,178 \\
\hline
\end{tabular}
\caption{\label{tab:KKIIc} Value of the four components of Feynman diagram IIc.}
\end{table}

The four components of IIc have abnormally high values \newline (-13, 34, -55 and 32) which surprisingly compensate for each other, resulting in -3,178, an order of magnitude lower. It is not possible to say anything more about the calculation of diagram IIc because the complete calculation was never published.

The authors indicate that \cite{KK}: \textbf{\textit{“The details of two independent calculations which were performed so as to provide some check of the final result are available from the authors”}}. That is, the authors affirm that the calculations were carried out independently by two teams of mathematicians who obtained the same result, as a guarantee that the calculations were correct.

\subsection{New experimental value}

Six years after the publication of the Karplus and Kroll's paper, Franken and Liebes \cite{Franken} published new and more precise experimental data that showed a very different value for the electron g-factor (1.001165). This value was higher than the Schwinger factor, so the value of the second coefficient calculated by Karplus and Kroll not only did not improve the Schwinger factor, but made it worse. With the new experimental data, the value of the second coefficient in the series should have been +0.7 instead of -2.973.

Karplus and Kroll admitted that it was not true that two independent calculations had been carried out, so it was possible that there were errors in the calculations. 

According to Kroll \cite{Kroll}: \textbf{\textit{“Karplus and I carried out the first major application of that program, to calculate the \linebreak fourth order magnetic moment, which calculation subsequently turned out to have some errors in it, which has been a perpetual source of embarrassment to me, but nevertheless the paper I believe was quite influential. (...) The errors were arithmetic (...) We had some internal checks but not nearly enough. (...) it was referred and published and was a famous paper and now it's an infamous paper.”}}.

The history of this correction is complex and confusing. We will now try to reconstruct this story from the published papers and quotes from its protagonists.

\markright{\ptepauthor. \pteptitle}
\section{The history of the correction}
\markright{\ptepauthor. \pteptitle}

\subsection{Petermann's numerical calculation}

Petermann was the first person to identify an error in the original calculation of Karplus and Kroll. He performed a numerical analysis of the five Feynman diagrams and he found that the solution of diagram IIc was clearly wrong, since its value was outside the limits. The rest of the diagrams were within limits \cite{Peter1}: \textbf{\textit{“The numerical results for the terms I, IIa, IIc, IId, IIe in the work by Karplus and Kroll have been checked by rigorous upper and lower bounds. Whereas every other term fell well between these bounds, agreement could not be obtained for diagram IIc. (...) The numerical value for this term has been found to satisfy IIc = -1.02 +/- 0.53.”}}.

Petermann published a second paper where he adjusted his calculations \cite{Peter2}: \textbf{\textit{”the diagram IIc is found to satisfy IIc = -0.60 +/- 0.11 in contradiction with the value -3.18 given by the previous authors.”}}.

Between the publication of these two papers,   Petermann communicated privately to Sommerfield the result of another calculation \cite{Sommer1}: \textbf{\textit{"Note added its proof. Petermann has \linebreak placed upper and lower bounds on the separate terms of Karplus and Kroll. He finds that their value for IIc does not lie within the appropriate bounds. Assuming the other terms to be correct, he concludes that the result is -0.53 +/- 0.37.”}}.

 Petermann worked for three months following a numerical methodology that allowed him to narrow the margin of error in diagram IIc. Surprisingly, 14 days after his third numerical calculation, he made an unexpected change in his methodology and published the exact analytical calculation, with no margins of error.

The articles published by Petermann on the calculation of the Feynman diagram IIc are summarized in the following table:

\renewcommand{\arraystretch}{1.3}
\begin{table}[ht]
\begin{tabular}{|c|c|c|p{2.7cm}|}
\hline
Date & IIc & Method & Publication \\
\hline
28/5 & -1.02 +/- 0.53 & Numerical & Nuclear Phys. 3 \\
\hline
1/7 & - 0.53 +/- 0.37 & Numerical & Phys. Rev. 107, \linebreak Note added in \linebreak proof. \linebreak Private comm.  \linebreak with \linebreak Sommerfield \\  
\hline
3/8 & -0.60 +/- 0.11 & Numerical & Nuclear Phys. 5 \\
\hline
17/8 & -0,564 & Analytical & Helvetica Physica Acta 30 \\
\hline
\end{tabular}
\caption{\label{tab:Petpapers} Petermann's publications.}
\end{table}

\subsection{Sommerfield and the Green’s functions}

After the publication of the new experimental value by \linebreak Franken and Liebes \cite{Franken}, Schwinger commissioned a 22-year-old student named Sommerfield to redo the Kroll and Karplus calculations. Schwinger proposed using his own method \linebreak based on Green's functions instead of using Feynman diagrams. 

According to Sommerfield's testimony \cite{Sommer100}: \textbf{\textit{"Julian assigned us three problems, one of which involved the anomalous magnetic moment (...). At my meeting with him, he suggested that I continue the calculation of the anomalous magnetic moment to the next fourth order (...). Schwinger wanted me to use the other method, while respecting gauge invariance at every step. Many years later Roy Glauber told me that the faculty was not entirely happy that a graduate student had been given such a problem."}}.

In May 1957, Sommerfield sent a two-page paper to the Physical Review Journal where he published his results\cite{Sommer2}: \textbf{\textit{“The fourth-order contribution to the moment is found to be -0.328 (..) Thus the result is 1.0011596.“}}. This new theoretical value of the electron g-factor was in good agreement with the new experimental value of Franken and Liebes.

As Schwinger states\cite{Mehra}: \textbf{\textit{“Interestingly enough, \linebreak although Feynman-Dyson methods were applied early [by Karplus and Kroll], the first correct higher order calculation was done by Sommerfield using [my] methods.”}}.

The second coefficient of the Dyson series calculated by Sommerfield consisted of four components, the same as the original result for Karplus and Kroll, but with very different values:

\medskip

[K\&K]
\begin{equation}
C_2 = -\frac{2687}{288} + \frac{125}{36} \pi ^2 - 9 \pi ^2 ln (2) + 28 \zeta(3) = -2,973
\end{equation}

\medskip

[Sommerfield]
\begin{equation}
C_2 = \frac{197}{144} + \frac{1}{12} \pi ^2 - \frac{1}{2} \pi ^2 ln (2) + \frac{3}{4} \zeta(3) = -0,328
\end{equation}

\renewcommand{\arraystretch}{1.3}
\begin{table}[ht]
\centering
\begin{tabular}{ |c|cccc|c|} 
\hline
&  Const. & $\pi^2$ & $\pi^2 \ ln(2)$ & $\zeta(3)$ & Total \\ 
\hline
K\&K & -9,329 & 34,269 & -61,569 & 33,656 & -2,973 \\
\hline
Pet. & 1,368 & 0,822 & -3,421 &  0,901 & -0,328 \\
\hline
Diff. & 10,697 & -33,447 & 58,148 & -32,754 & 2,645 \\
\hline
\end{tabular}
\caption{\label{tab:compC2} Comparative components of $C_2$.}
\end{table}

Sommerfield's paper does not include the calculations \linebreak performed, but the author states that \cite{Sommer1}: \textbf{\textit{“The present calculation has been checked several times and all of the auxiliary integrals have been done in at least two different ways.”}}. As a guarantee that the calculations were correct. 

In 1958 Sommerfield published his g-factor calculations in the Annals of Physics \cite{Sommer2} as part of his doctoral thesis. If we analyze his extensive 32-page paper, we verify that he used Green's functions instead of Feynman diagrams. For this reason, the calculation of the enigmatic Feynman diagram IIc does not appear in this paper.

In the third volume of "Particles, Sources, and Fields" published in 1989 \cite{Schwinger}, Schwinger devoted more than 60 pages to a detailed calculation of the second coefficient of Dyson series getting exactly the same result. But, once again, using Green's functions instead of Feynman diagrams.

In his 1957 paper Sommerfield also states that \cite{Sommer1}: \textbf{\textit{“The discrepancy has been traced to the term I y IIc of Karplus and Kroll.”}}. This statement about the origin of the error cannot be deduced from Sommerfield's calculations, \linebreak since he used Green's functions instead of Feynman \linebreak diagrams. So Sommerfield had to receive this information from other sources (Petermann, Karplus or Kroll).  
 
\subsection{Petermann's definitive correction}

The definitive solution to the problem was presented in 1957 by Petermann in a paper published in the 
Swiss journal Helvetica Physica Acta \cite{Peter}. Although the paper was signed by a single author, actually the result was obtained by consensus between the results of the Petermann's numerical analysis, the Sommerfield calculation of $C_2$ using Green's functions and the correction of the Feynman diagrams carried out by Kroll himself. Petermann acknowledges that the result was obtained by consensus \cite{Peter}: \textbf{\textit{“The new fourth order correction given here is in agreement with: (a) The upper and lower bounds given by the author. (b) A calculation using a different method, performed by C. Sommerfield. (c) A recalculation done by N. M. Kroll and collaborators.”}}.

The article was signed by a single author due to an internal conflict between the researchers. As Sommerfied recalls \cite{Sommer100}: \textbf{\textit{"In the meantime Schwingerian Paul Martin had gone to the Niels Bohr Institute in Copenhagen and had spoken to Andre Petermann, a postdoc with the Swedish theoretician Gunnar Kallen. Martin told Petermann about my work (...)  In the end, however, after both of our calculations were completely finished they were in agreement with each other but not with Karplus and Kroll. We agreed to cite each other's work when published. However, Schwinger and Kallen had had a somewhat acrimonious discussion (...) and Kallen had forbidden Petermann to mention my work. Petermann's apology to me was profuse."}}. 

The Petermann's final result for the electron g-factor was identical to the  Sommerfield's result published three \linebreak months earlier. 
\begin{equation}
C_2 = \frac{197}{144} + \frac{1}{12} \pi ^2 - \frac{1}{2} \pi ^2 ln (2) + \frac{3}{4} \zeta(3) = -0,328
\end{equation}

In the paper, Petermann states that: \textbf{\textit{“We have performed an analytic evaluation of the five independent diagrams \linebreak contributing to this moment in fourth order. The results are the following (I = -0.467, IIa = 0.778, IIc = -0.564, IId = -0.090, IIe = 0.016, Total = -0.328). Compared with the values in their original paper by Karplus and Kroll, one can see that two terms were in error: I differs by 0.031 and IIc differs by 2.614.”}}. 

\renewcommand{\arraystretch}{1.3}
\begin{table}[h]
\centering
\begin{tabular}{ |ccccc|c|} 
\hline
I & IIa & IIc & IId & IIe & Total \\
\hline
-0.467 & 0.778 & -0.564 & -0.090 & 0.016 & -0.328 \\
\hline
\end{tabular}
\caption{\label{tab:Pet} Corrected values of the five Feynman diagrams.}
\end{table}

Comparing the results of the calculations of the Feynman IIc diagram carried out by Karplus and Kroll with the Petermann calculations we observe the following:

\medskip

[K\&K]
\begin{equation}
  II_c = -\frac{323}{24} + \frac{31}{9} \pi ^2 -  \frac{49}{6} \pi ^2 ln (2) +  \frac{107}{4} \zeta(3) 
\end{equation}

[Petermann]
\begin{equation}
II_c = -\frac{67}{24} +  \frac{1}{18} \pi ^2 +  \frac{1}{3} \pi ^2 ln (2) -  \frac{1}{2} \zeta(3) 
\end{equation}

\medskip

The calculation of each of the four factors in diagram IIc is shown in the following table:

\renewcommand{\arraystretch}{1.3}
\begin{table}[h]
\centering
\begin{tabular}{ |c|cccc|c|} 
\hline
&  Const. & $\pi^2$ & $\pi^2 \ ln(2)$ & $\zeta(3)$ & Total \\ 
\hline
K\&K & -13,458 & 33,995 & -55,868 & 32,153 & -3,178 \\
\hline
Pet. & -2,791 & 0,548 & 2,280 & -0,601 & -0,564 \\
\hline
Diff. & 10,667 & -33,447 & 58,148 & -32,754 & 2,614 \\
\hline
\end{tabular}
\caption{\label{tab:compIIc} Comparative components of Feynman diagram IIc.}
\end{table}

The corrections are huge, one or two orders of magnitude for each component of diagram IIc. We cannot know the origin of these discrepancies because the correction calculations were also not published.

\markright{\ptepauthor. \pteptitle}
\section{Summary}
\markright{\ptepauthor. \pteptitle}

The calculation of the Feynman diagram IIc can be considered the most important calculation in the history of modern physics. However, the history of this calculation is surrounded by  big errors and inexplicable coincidences.

\begin{itemize}

\item The original calculation of the Feynman diagram  IIc published in 1950 was completely wrong. 

\item Karplus and Kroll stated that the calculation had been performed by two teams independently. This statement was made to give guarantees about the validity of the calculations, and yet it turned out to be false.

\item Despite having published a completely wrong result, the prestige of Karplus and Kroll was not affected at all. On the contrary, both enjoyed brilliant careers full of awards and recognition for their professional achievements.

\item The Karplus and Kroll miscalculation was consistent with the experimental value previously published by \linebreak Gardner and Purcell, even though that experimental \linebreak value was also wrong.

\item The error in the calculation was not reported until seven years after its publication. 

\item The error in the calculation was detected just when a new experimental value was published by Franken and Liebes. The corrected theoretical value also coincided with the new experimental value.

\item Neither the original calculation of the Feynman diagram IIc nor its subsequent correction has been published to date.

\end{itemize}

\begin{flushright}\footnotesize
1 September 2020
\end{flushright}

\vspace*{-6pt}
\centerline{\rule{72pt}{0.4pt}}
}

\end{document}